\input{aipcheck}

\documentclass[
    ,final            
  ]
  {aipproc}

\layoutstyle{6x9}

\usepackage{amssymb,amsmath,array}
\usepackage{wrapfig,rotating}
\usepackage{xspace}
\setcitestyle{numbers}

\begin{document}

\newcommand{\dzero}     {D\O\xspace}
\newcommand{\wplus}     {$W+$jets\xspace}
\newcommand{\zplus}     {$Z+$jets\xspace}
\newcommand{\muplus}    {$\mu +$jets\xspace}
\newcommand{\eplus}     {$e +$jets\xspace}
\newcommand{\ljets}     {$l +$jets\xspace}


FERMILAB-CONF-11-390-PPD

\title{Measurement of the Properties of the top Quark at \dzero}


\author{Andreas W. Jung~~(for the \dzero~collaboration)\\}{
        address={Fermi National Accelerator Laboratory (Fermilab)\\
        E-mail: ajung@fnal.gov}
}



\begin{abstract}
Different measurements of the properties of the $top$ quark using up to $5.4~\mathrm{fb^{-1}}$ collected with the
\dzero~detector at the Fermilab Tevatron collider are presented. The top mass is obtained from a study of dilepton and lepton+jets final states, while the width is obtained from a combination of the measurements of the single $top$ production via $t$-channel exchange and the determination of the $t\rightarrow Wb$ branching ratio. Furthermore the measurement of the helicity of the $W$ boson from $top$ quark decays, a measurement of $t\bar{t}$ spin correlations and a measurement of the jet pull (color flow) in $t\bar{t}$ events are presented.
\end{abstract}

\maketitle


\section{Introduction}
The $top$ quark is the heaviest known elementary particle and was discovered at the Tevatron collider in 
1995 \cite{top_disc1, top_disc2} at a mass of around $175~\mathrm{GeV}$. The dominant production channel at the 
Tevatron in $p\bar{p}$ collisions is via $q\bar{q}$ annihilation with 85\% as opposed to gluon-gluon fusion which contributes only 15\%. 
Many details of the analysis method and final results refer to the decay channel of the $t\bar{t}$ pair, i.e. if 
either none, one or both of the $W$ bosons (stemming from decay of the $top$ quarks) decay semi-leptonically. The dominant decay channel is the "all jets" channel, followed by the "lepton+jets" channel (\ljets) and the "dilepton" channel. The reason for these categories is the drastic change in amount of background in each of the channels with the "all jets" being worst. The 
\ljets~channel is a good compromise between signal and background contribution whilst having high event statistics, whereas the "dilepton" channel has very small backgrounds but also reduced event statistics due to the small branching fractions.

\section{Measurement of top properties}
In the following recent measurements of the properties of the $top$ quark in the \ljets~and dilepton decay 
channel at \dzero~are summarized. First common analysis issues as well as systematics are briefly discussed. 
The main background contribution in the \ljets~decay channel originates from \wplus~production whereas the 
dilepton decay channel suffers most from contributions from \zplus~production.\\
The dominant systematic error sources for measurements of top properties are the uncertainty of the 
jet energy scale (JES) and resolution, modeling of the signal and production including hadronization, 
higher orders and also the choice of the parton density distribution functions. Depending on the analysis also
the systematic error of the b-tagging efficiency plays a major role.

\subsubsection{Top Quark Mass}
\begin{figure}[ht]
    \begin{minipage}[c]{0.475\columnwidth}
    \centerline{\includegraphics[width=0.9\columnwidth]{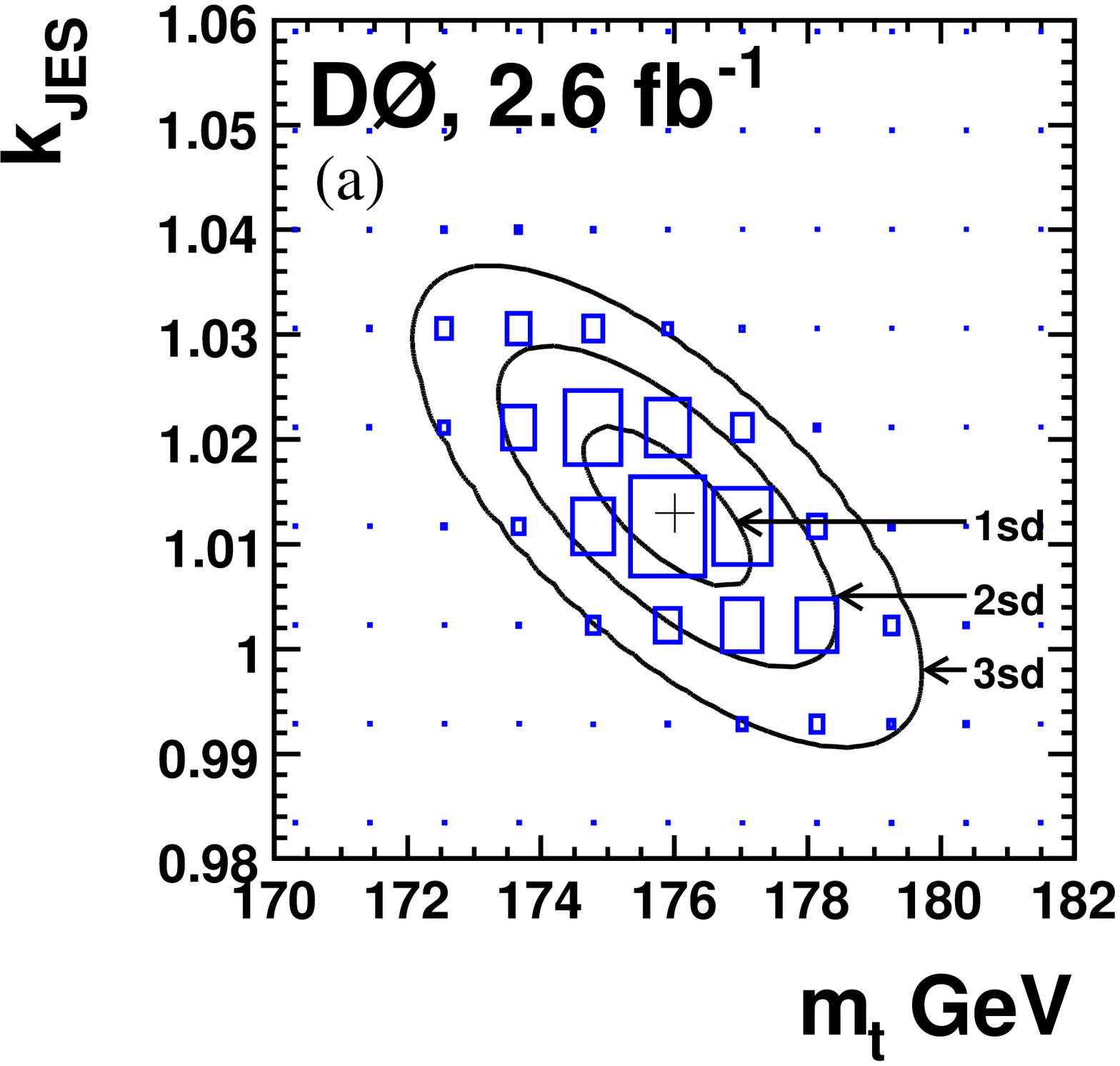}}
     \end{minipage}
     \hspace{.05\linewidth}
     \begin{minipage}[c]{0.475\columnwidth}
    \centerline{\includegraphics[width=0.95\columnwidth]{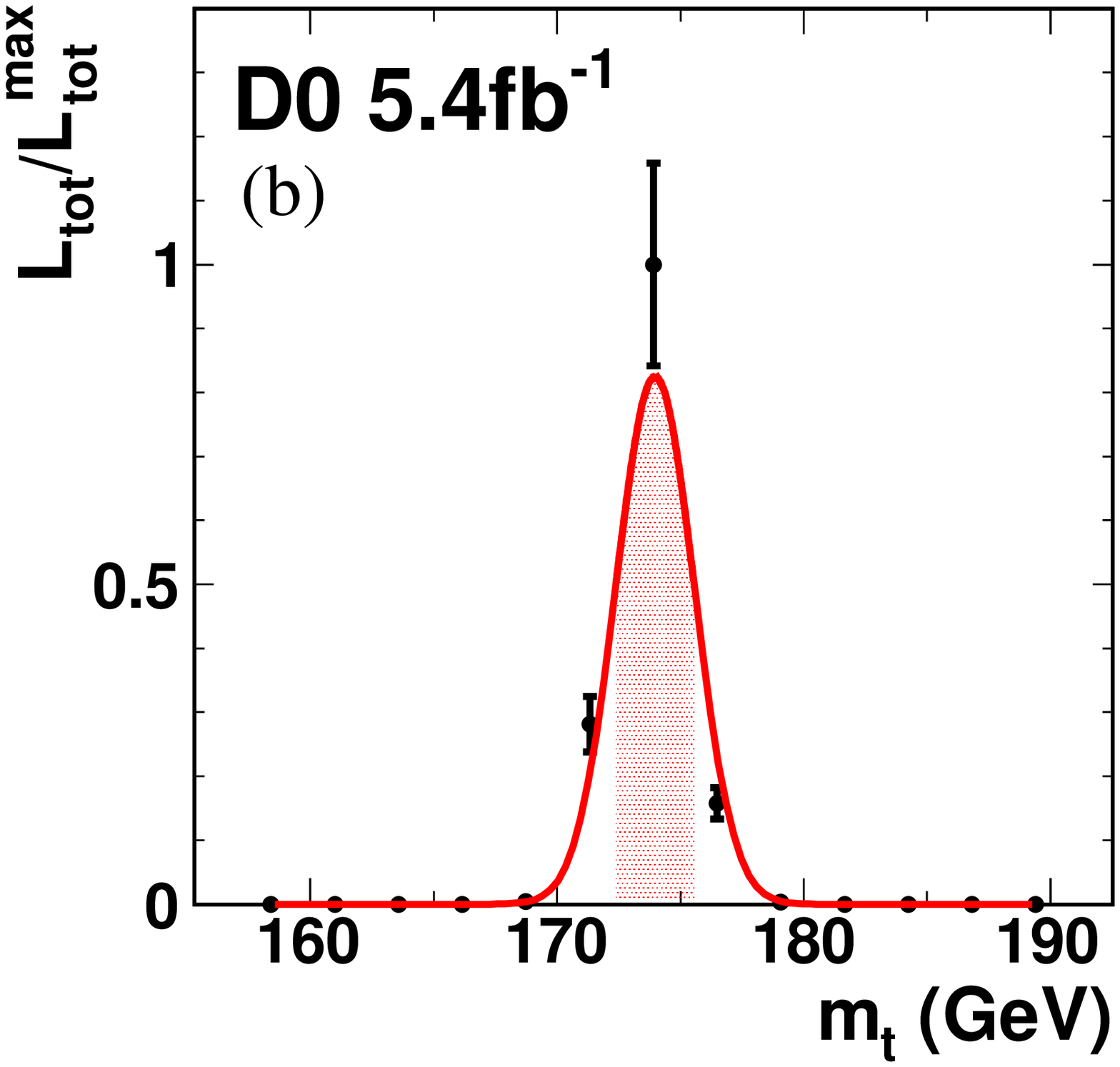}}
     \end{minipage}
  \caption{\label{fig:mass} Fitted contours of equal probability for the two-dimensional likelihood as a function
 of $k_{\mathrm{JES}}$ against $m_t$ (a) with the best fit and 1, 2 and 3 standard deviation contours for the 
\ljets~decay channel. The calibrated and normalized likelihood for data (b) as a function of $m_t$ with 
best estimate as well as 68\% confidence level region marked by the shaded area for the dilepton decay channel.}
 \end{figure}

The first presented measurement is the latest $top$ quark mass measurement by \dzero. It uses the so-called 
matrix element method (ME) which calculates an event probability density from differential cross sections and
 detector resolutions. The transfer-function relates the probability density of measured quantities to the partonic
 quantities. As one of the $W$ bosons decays hadronically a constraint on the $W$ mass can be used to fit the jet 
energy scale in-situ. Figure \ref{fig:mass}(a) shows the JES against the $top$ mass. The currently most 
precise \dzero~mass measurement uses $3.6~\mathrm{fb^{-1}}$ and yields a mass of 
$m_{t} = 174.9 \pm 0.8 (\mathrm{stat.}) \pm 1.2 (\mathrm{sys.\,+\,JES})~\mathrm{GeV}$ \cite{mass}.\\
The $top$ mass is also measured from events in the dilepton decay channel using the matrix element method. This sample
 is limited in statistics due to the small BR but has very low backgrounds. There is no in-situ JES correction possible
 as both $top$ quarks decay leptonically. Figure \ref{fig:mass}(b) shows the calibrated and normalized likelihood for 
data using $5.4~\mathrm{fb^{-1}}$. The measurement yields $m_{t} = 173.6 \pm 1.8 (\mathrm{stat.}) \pm 2.5 (\mathrm{sys.})~\mathrm{GeV}$ \cite{mass_dilepton}.
\begin{figure}[ht]
    \begin{minipage}[c]{0.475\columnwidth}
    \centerline{\includegraphics[width=0.9\columnwidth]{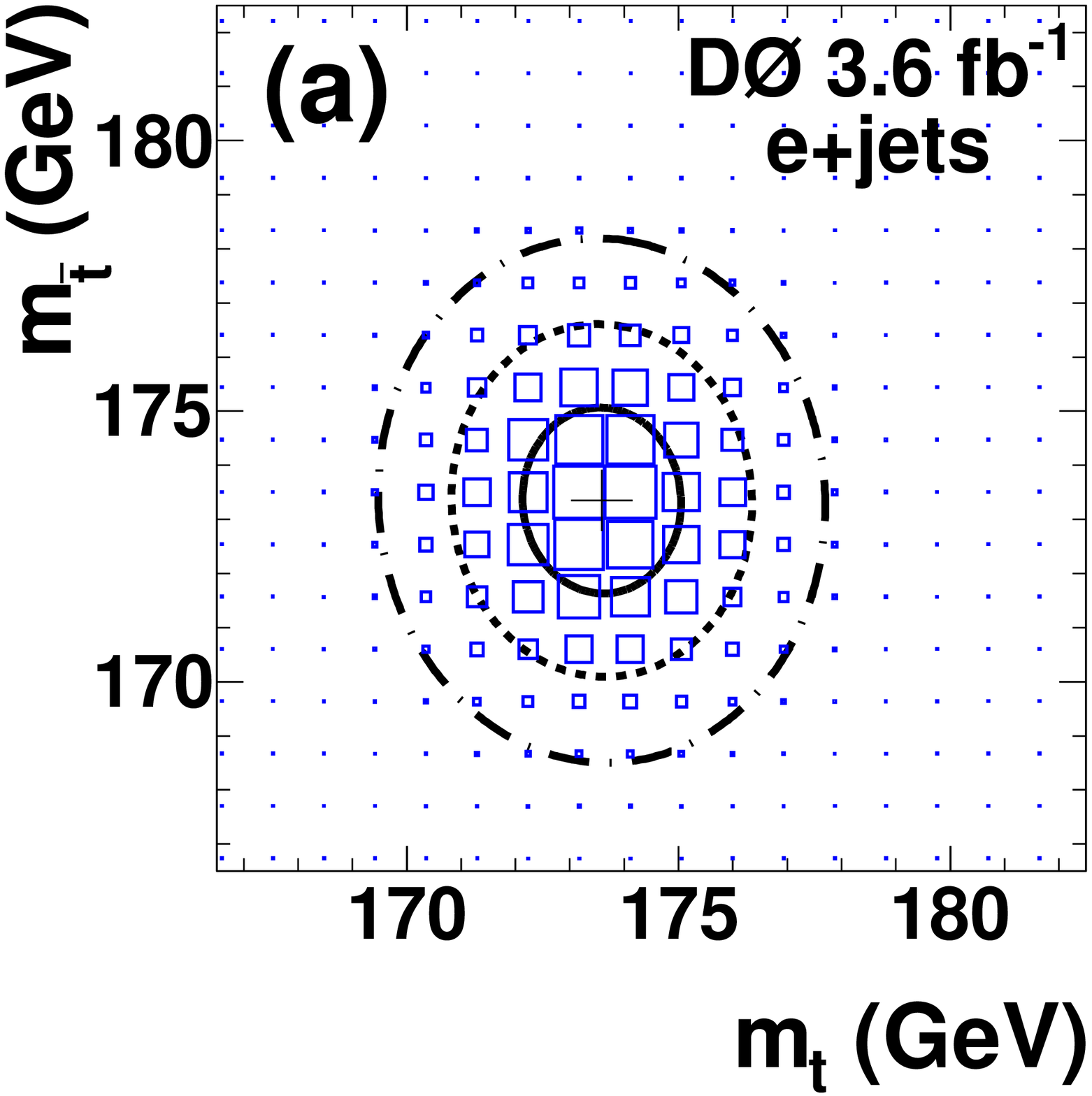}}
     \end{minipage}
     \hspace{.05\linewidth}
     \begin{minipage}[c]{0.475\columnwidth}
    \centerline{\includegraphics[width=0.9\columnwidth]{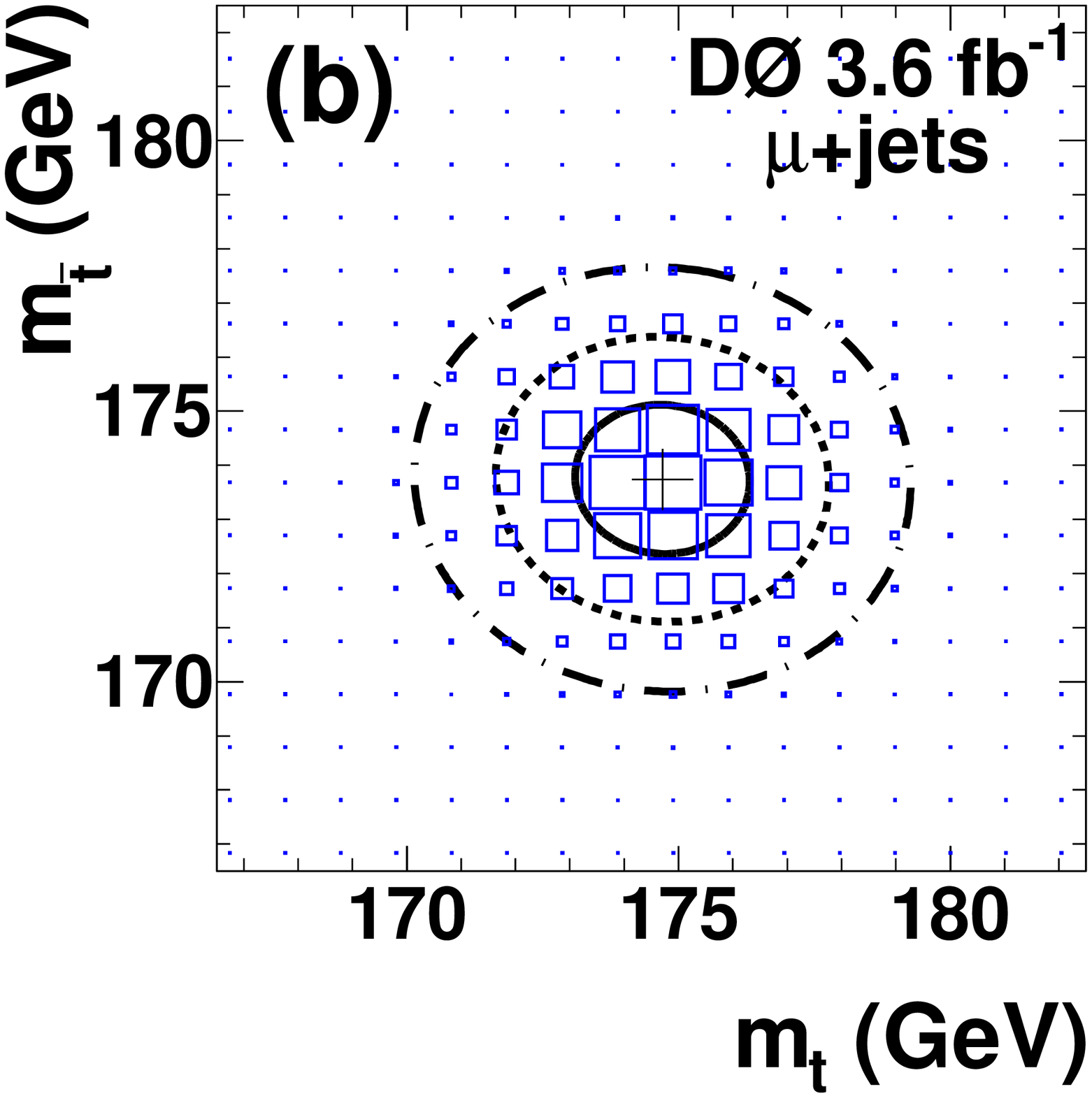}}
     \end{minipage}
  \caption{\label{fig:deltaM} Combined likelihoods of the RunIIa and RunIIb measurements as functions of $m_t$ and 
$m_{\bar{t}}$ in data for the (a) \eplus~and (b) \muplus~channel. The solid, dashed, and dash-dotted lines 
represent the 1, 2, and 3 SD contours.}
 \end{figure}
Furthermore the very same event selection as for the $top$ mass measurement in the \ljets~channel has been used to 
measure the mass difference $\Delta M_{t}$ between $top$ and anti-$top$ quarks (see Figure \ref{fig:deltaM}). The
mass difference is predicted to be $0$ in the standard model (SM). The analysis yields 
$\Delta M_{t} = 0.8 \pm 1.8 (\mathrm{stat.}) \pm 0.8 (\mathrm{sys.})~\mathrm{GeV}$ 
and does not show any indication of a $top$ anti-$top$ mass difference \cite{deltaM}.\\

\subsubsection{Top Decay Width}
\begin{figure}[ht]
    \begin{minipage}[c]{0.475\columnwidth}
    \centerline{\includegraphics[width=0.9\columnwidth]{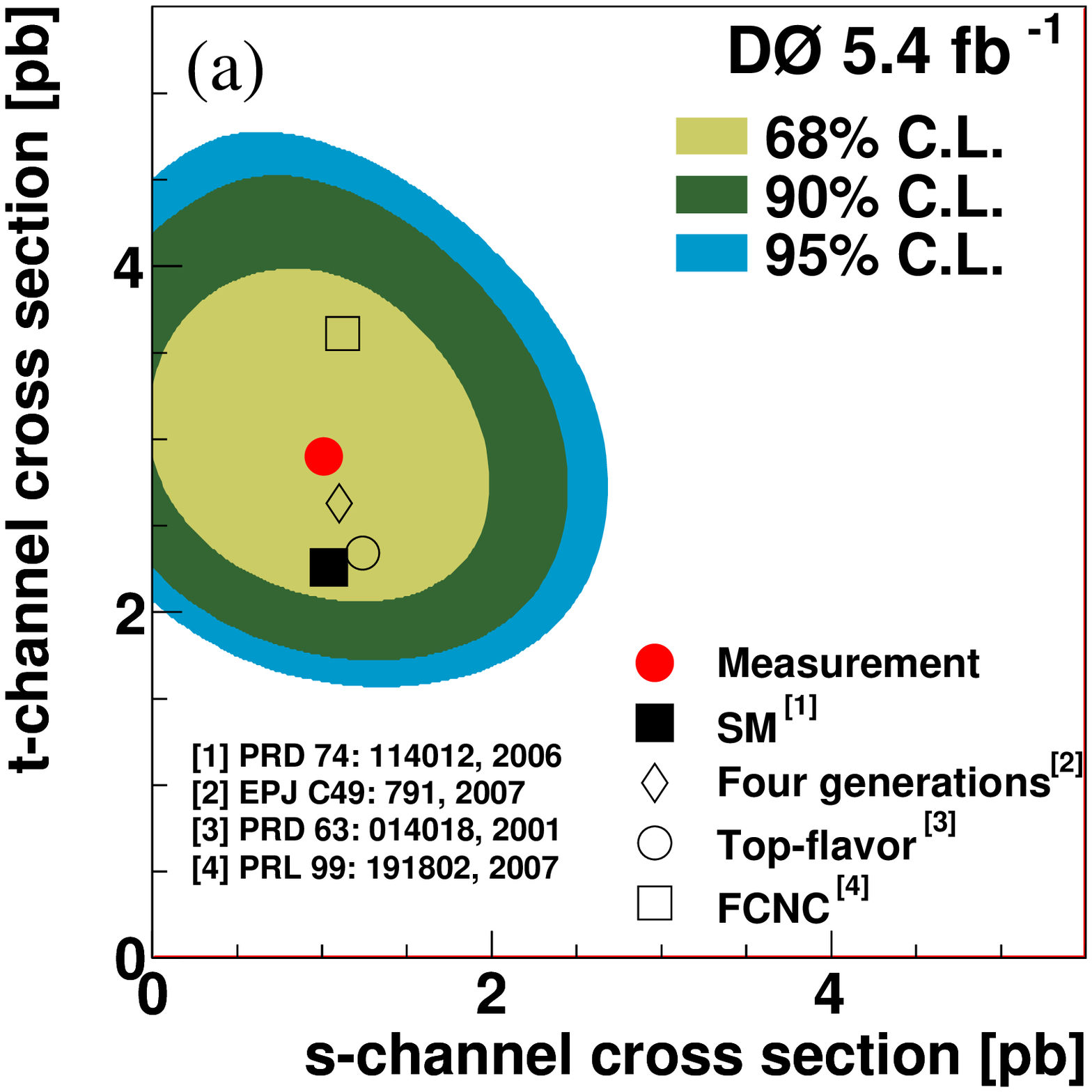}}
     \end{minipage}
     \hspace{.05\linewidth}
     \begin{minipage}[c]{0.475\columnwidth}
    \centerline{\includegraphics[width=0.85\columnwidth]{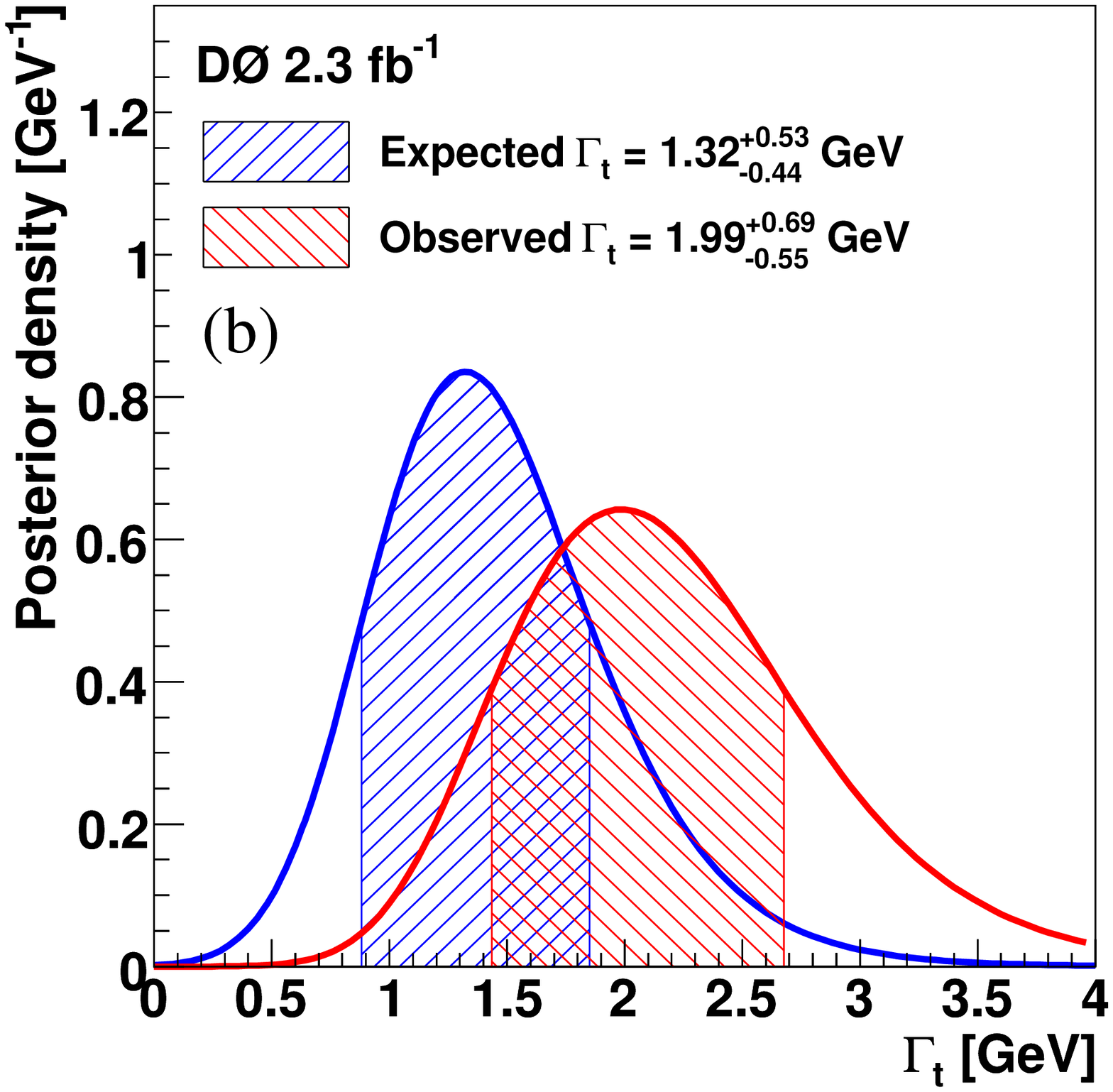}}
     \end{minipage}
  \caption{\label{fig:width} The result of the measurement of the single $top$ production cross 
section in the $s$- and $t$-channel (a). The latter is used to determine the total width of the $top$ quark (b).}
 \end{figure}

The indirect measurement of the decay width of the $top$ quark uses two preceding measurements, that is the $t$-channel single $top$ production cross section and the measurement of the ratio of the branching fractions 
$t \rightarrow Wq$ \cite{r_measure}. Using theoretical calculations and the measured $t$-channel cross section (see Figure \ref{fig:width}(a)) the partial decay width $\Gamma(t\rightarrow Wb)$ can be determined:
\begin{equation}
\Gamma(t\rightarrow Wb) = \sigma (t\mathrm{-channel}) \frac{\Gamma(t\rightarrow Wb)_{\mathrm{SM}}}{\sigma (t\mathrm{-channel})_{\mathrm{SM}}}~.
\end{equation}
The total decay width $\Gamma_{t}$ of the $top$ quark is given by:
\begin{equation}
\Gamma_{t} = \frac{\Gamma(t\rightarrow Wb)}{B(t\rightarrow Wb)}~.
\end{equation}
By correcting the partial decay width with the result of the measurement of the ratio of the branching fractions 
$t \rightarrow Wq$: $\mathrm{BR}(t \rightarrow Wb)= 0.97^{+0.09}_{-0.08}$ the total width of the $top$ quark is measured to be $1.99^{+0.69}_{-0.55}~\mathrm{GeV}$ (see Figure \ref{fig:width}(b)) \cite{width}. This is the most precise indirect determination of the total decay width of the $top$ quark. The value translates into
 a lifetime of $\left (3.3^{+1.3}_{-0.9} \right ) \times 10^{-25}~\mathrm{s}$.

\subsubsection{W helicity}
\begin{figure}[ht]
    \begin{minipage}[c]{0.475\columnwidth}
    \centerline{\includegraphics[width=0.9\columnwidth]{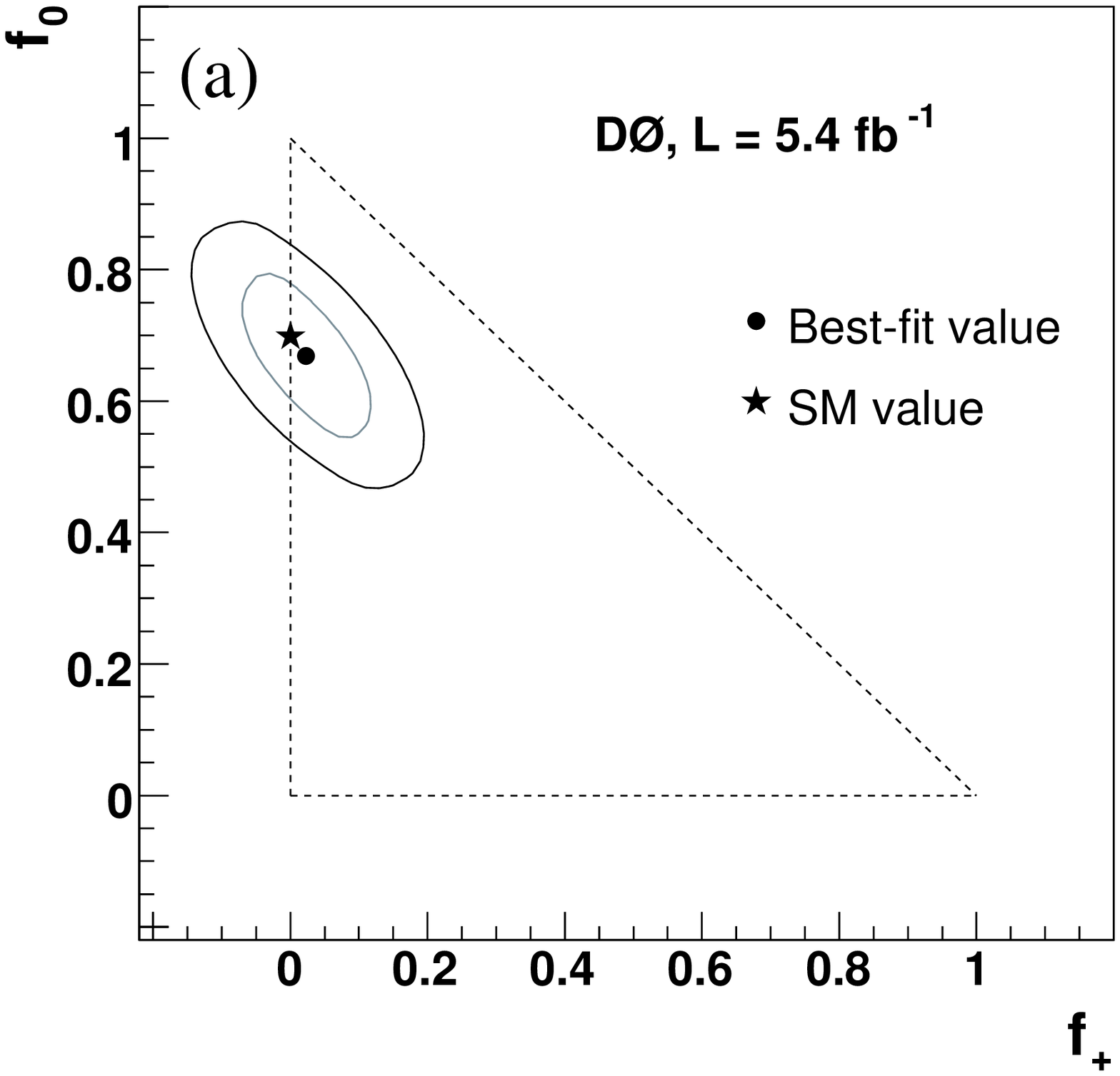}}
     \end{minipage}
     \hspace{.05\linewidth}
     \begin{minipage}[c]{0.475\columnwidth}
    \centerline{\includegraphics[width=0.925\columnwidth]{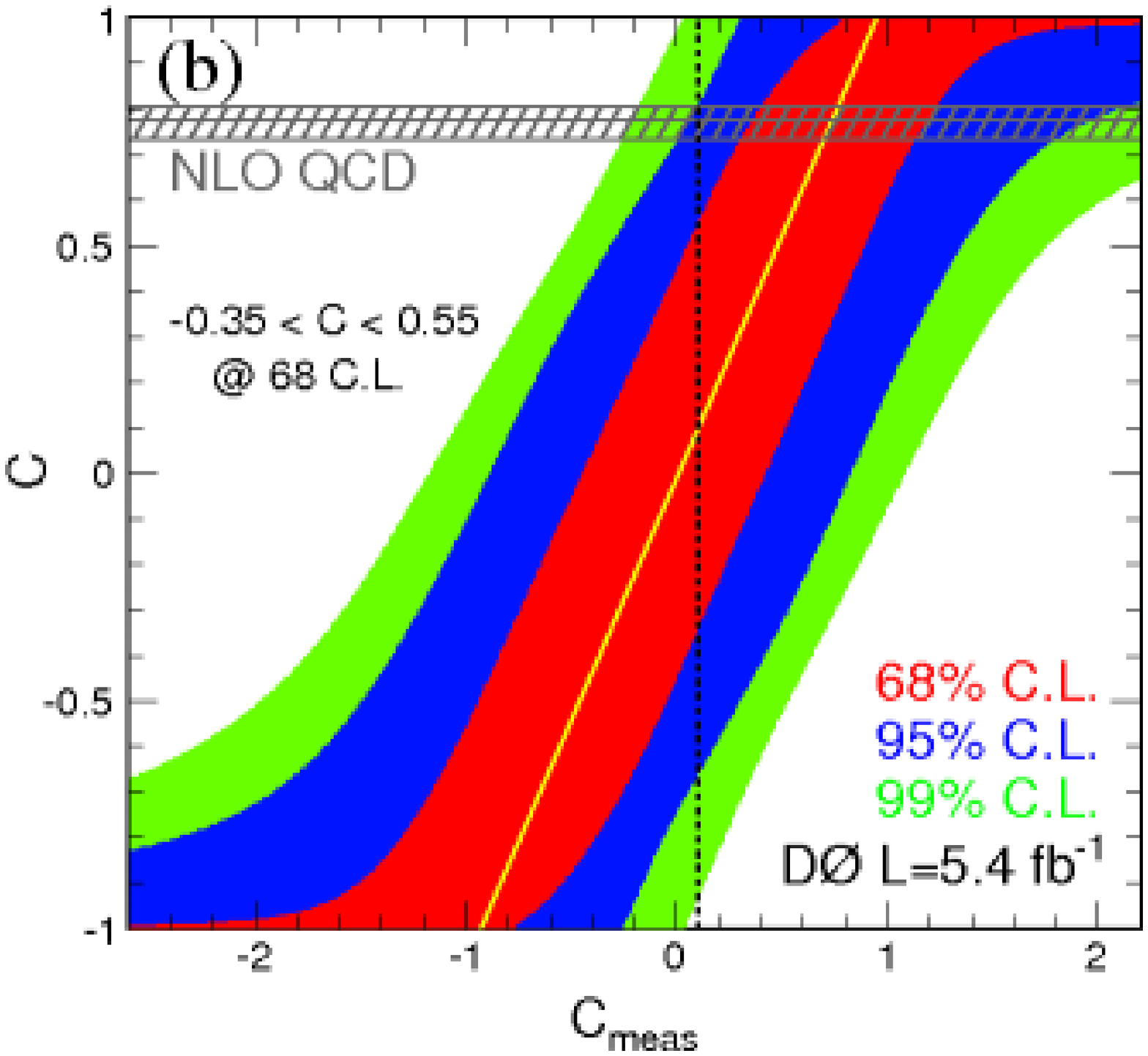}}
     \end{minipage}
  \caption{\label{fig:helicity_spin} Result of the W boson helicity fit for the combined RunIIa and RunIIb data sample (a). 
The ellipses indicate the 68\% and 95\% C.L. contours, the dot shows the best-fit value, the triangle corresponds to the physically
 allowed region where $f_0 + f_+ \le 1$, and the star marks the expectation from the SM. (b) shows the 68\% (inner), 
95\% (middle), and 99\% (outer) C.L. bands of $C$ as a function of $C_{\mathrm{meas}}$ from likelihood fits to MC events for 
all channels combined. The vertical dashed black line depicts the measured value $C_{\mathrm{meas}} = 0.10$. The horizontal 
band indicates the NLO QCD prediction of $C = 0.777^{+0.027}_{-0.042}$.}
 \end{figure}

The $V-A$ nature of the weak force gives rise to a prediction for the polarization of $W$ bosons originating from $top$ 
quark decays. The standard model (SM) predictions for left-handed, longitudinal and right-handed polarizations are: 
$f_0=0.698,~f_-=0.301$ and $f_+=O(10^{-4})$ \cite{whelicity_calc}. The measurement uses $5.4~\mathrm{fb^{-1}}$ and the best fit values (see Figure \ref{fig:helicity_spin}(a)) are $f_0 = 0.669 \pm 0.078 (\mathrm{stat.}) \pm 0.065 (\mathrm{sys.})$ and 
$F_+ = 0.023 \pm 0.041 (\mathrm{stat.}) \pm 0.034 (\mathrm{sys.})$ \cite{w_helicity} which is consistent with the standard model.

\subsubsection{$t\bar{t}$ Spin correlations}
A measurement that is complementary at the Tevatron as compared to the LHC because of the initial state is the measurement 
of spin correlations in $t\bar{t}$ events. The very short lifetime of the $top$ quark prevents spins from being affected by 
the fragmentation process. In other words the spin information is preserved and can affect the decay products, thus $t\bar{t}$ spin correlations are detectable. The next-to-leading (NLO) order QCD calculation predicts a correlation strength of $C=0.78^{+0.03}_{-0.04}$ in the beam basis \cite{correlations_calc}. Figure \ref{fig:helicity_spin}(b) shows the result of the measurement using dilepton events in $5.4~\mathrm{fb^{-1}}$. A correlation strength of $C=0.10 \pm 0.45~(\mathrm{stat.+sys.})$ has been measured \cite{spin}. A new measurement uses events that decay in the dilepton decay channel and determines the fraction of $t\bar{t}$ events with spin correlation using a matrix element approach \cite{spin_me}. A correlation strength of $C=0.57 \pm 0.31~(\mathrm{stat.+sys.})$ has been measured.

\subsubsection{Forward-backward Asymmetry $A_{fb}$}
NLO QCD gives rise to a forward-backward asymmetry in $t\bar{t}$ events due to interference terms. Because of the initial state the measurement is different between $q\bar{q}$ (Tevatron) and $qq$ (LHC). The asymmetry is larger at the Tevatron as the initial state is $q\bar{q}$ dominated as opposed to the LHC where it is $gg$ dominated.\\ 
The difference in rapidity $\Delta y = (y_{t} - y_{\bar{t}})$ between the two $top$ quarks is a measure of the asymmetry $A_{fb}$ given by:
\begin{equation}
A_{fb} = \frac{N^{\Delta y>0} - N^{\Delta y<0}}{N^{\Delta y>0} + N^{\Delta y<0}}~.
\end{equation}
Using $4.3~\mathrm{fb^{-1}}$ a forward-backward asymmetry of $A_{fb} = 8 \pm 4 (\mathrm{stat.}) \pm 1 (\mathrm{sys.}) \%$ is
 measured compared to the MC@NLO prediction of $(2.4 \pm 0.7)$\%. Recently this has been updated \cite{updated_afb}.

\subsubsection{Color Flow}
\begin{figure}[ht]
    \begin{minipage}[c]{0.475\columnwidth}
    \centerline{\includegraphics[width=0.925\columnwidth]{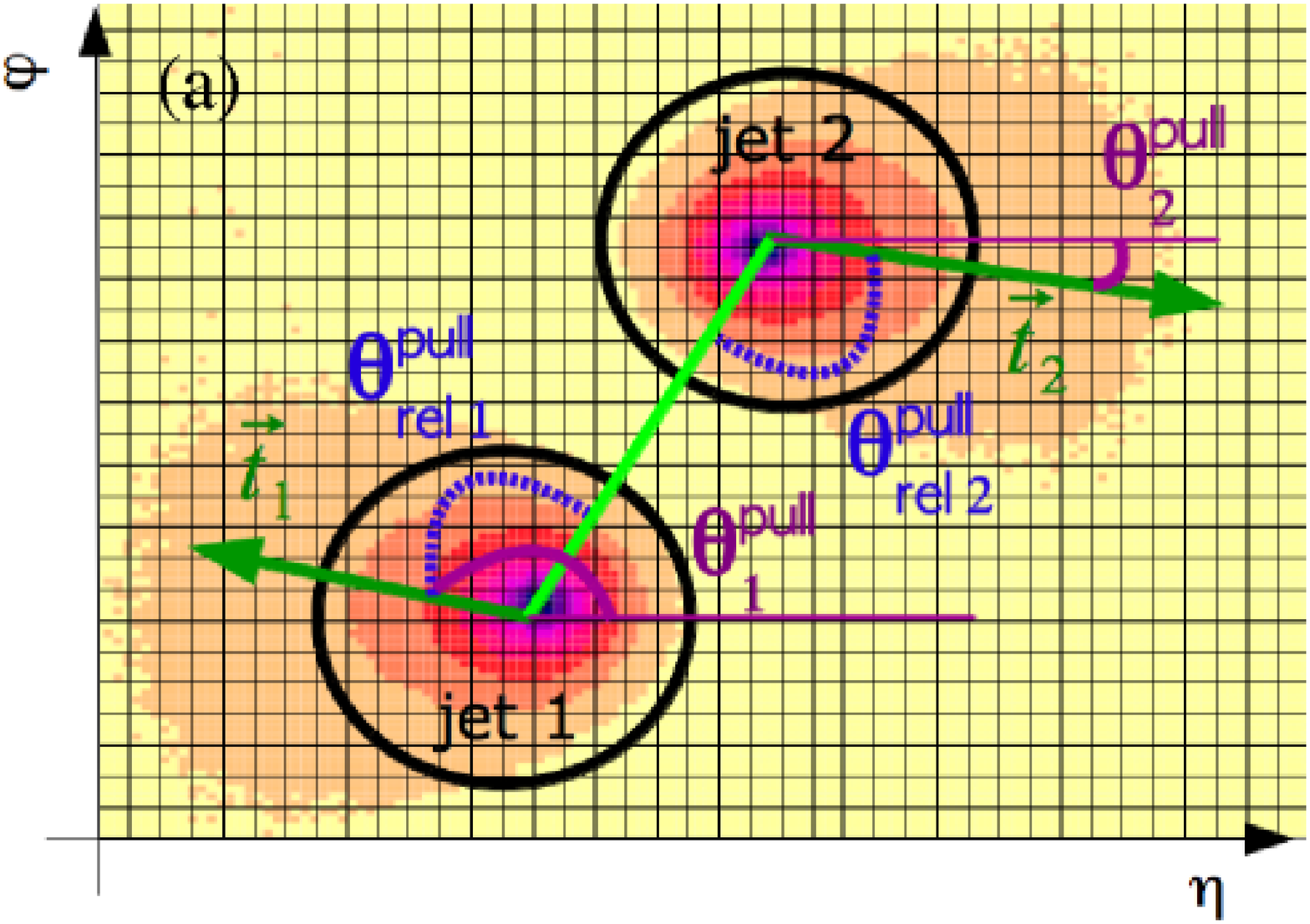}}
     \end{minipage}
     \hspace{.05\linewidth}
     \begin{minipage}[c]{0.475\columnwidth}
    \centerline{\includegraphics[width=0.9\columnwidth]{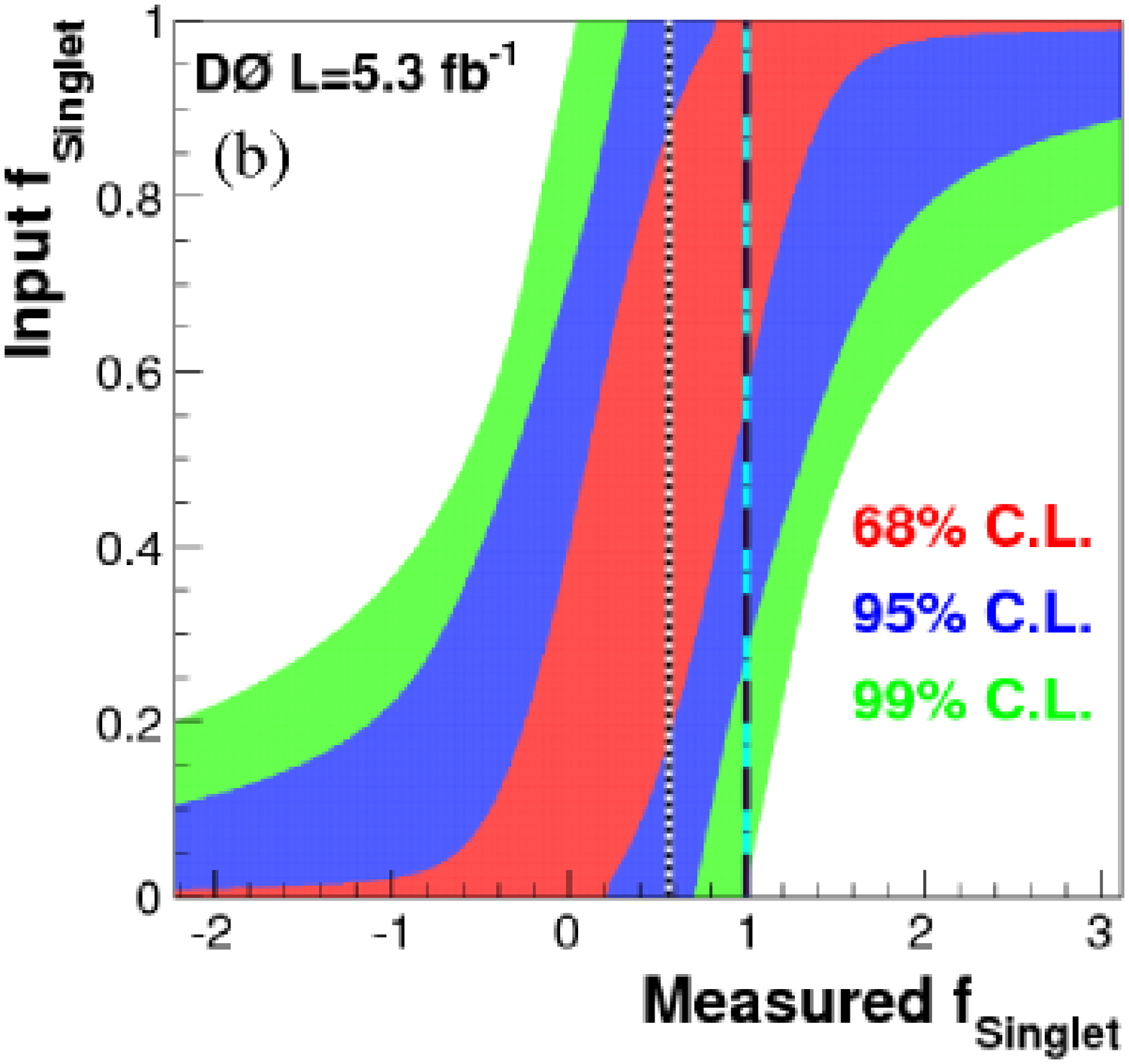}}
     \end{minipage}
  \caption{\label{fig:colorFlow} Schematic drawing (a) showing two jets in the $\eta-\phi$ plane, and the reconstruction of the jet pull vectors 
$(\,\vec{t}\,)$, jet pull angles ($\theta_{\mathrm{pull}}$), and relative jet pull angles ($\theta_{\mathrm{pull,\,rel}}$). (b) shows the expected C.L. 
bands for $f_{\mathrm{Singlet}}$. The measured value is shown on the horizontal axis, and the input value on the vertical axis.
 The wide-dashed line shows the expected value and the black-white fine-dashed line indicates the measured value of $f_{\mathrm{Singlet}}$.
}
 \end{figure}

$t\bar{t}$ events provide a relatively clean source for hadronically decaying $W$ bosons and can therefore be used to study the color flow in an event. Partons carry a color charge which provides additional information about the event. A color singlet like the Higgs decaying to $b\bar{b}$ will not be color-connected to the beam as the two $b$ quarks stem from a color neutral object. A color octet like a gluon decaying to $b\bar{b}$ will be color-connected to the beam as the gluon carries a color by itself. Figure \ref{fig:colorFlow}(a) shows a schematic drawing of this useful analysis technique using jet pull to study the color flow.

To show the validity of the method two different hypotheses for the hadronically decaying $W$ bosons are assumed: SM $t\bar{t}$ with a color-singlet hadronically decaying $W$ boson or $t\bar{t}$ with a color-octet hadronically decaying $'W'$ boson. $f_{\mathrm{singlet}}$ gives the fraction of events that originate from color-singlet $W$ boson decays with a SM prediction of $f_{\mathrm{singlet}} = 1$. Figure \ref{fig:colorFlow}(b) shows the limit for $f_{\mathrm{singlet}}$ derived using $5.4~\mathrm{fb^{-1}}$: $f_{\mathrm{singlet}} = 0.56 \pm 0.38 (\mathrm{stat.+sys.}) \pm 0.19 (\mathrm{MC stat.})$ \cite{flow}.

\section{Conclusion}
A wealth of measurements of properties of the $top$ quark at \dzero~have been discussed showing the great performance of the Tevatron and the \dzero detector. All results are consistent with the standard model expectations. The final \dzero~data sample will have $2-3$ times the presented statistics allowing for new \& more precise results in the future.




\end{document}